\documentclass[10pt,twocolumn]{article}

\usepackage[margin=0.75in]{geometry}
\usepackage{microtype}
\usepackage{booktabs}
\usepackage{multirow}
\usepackage{enumitem}
\usepackage{xcolor}
\usepackage{amsmath}
\usepackage{amssymb}
\usepackage{graphicx}
\usepackage{stfloats}
\usepackage{url}
\usepackage{xspace}
\usepackage[hidelinks]{hyperref}

\setlist[itemize]{leftmargin=*,topsep=2pt,itemsep=2pt}
\setlist[enumerate]{leftmargin=*,topsep=2pt,itemsep=2pt}
\newcommand{\sys}{MetaInfer\xspace}

\title{Automatic Model-Hardware Co-Adaptation for Heterogeneous AI Accelerators}

\author{
Tian Chen\\
Pazhou Laboratory (Huangpu)
\and
Mingheng Mi\\
Pazhou Laboratory (Huangpu)
\and
Pu Wang\\
Pazhou Laboratory (Huangpu)
\and
Chen Hu\\
Unaffiliated\\
\texttt{hclearner2@gmail.com}
\and
Hai Zhang\\
Pazhou Laboratory (Huangpu)
}
\date{}

\begin{document}
\twocolumn[
\begin{@twocolumnfalse}
\maketitle

\begin{abstract}
Large language models now evolve faster than production inference systems can be ported and optimized. New releases change attention, MoE routing, quantization formats, KV-cache layout, and parallel execution patterns, while deployed accelerator fleets remain heterogeneous across hardware generations, framework forks, operator libraries, compiler backends, and communication runtimes. Serving a new model on existing hardware is therefore a model-framework-kernel-hardware co-adaptation problem.

We present \sys, an LLM-agent system that formulates inference adaptation as route search over a costed execution-adaptation graph. The graph connects model semantics, framework dispatch, kernel choices, hardware capabilities, runtime evidence, and serving objectives. \sys constructs and updates this graph during execution, restores missing or blocked routes through patches, and reduces route cost through staged validation and end-to-end profiling. Three real episodes--DeepSeek V4 Flash on NVIDIA A800, GLM 5.3 Flash on NVIDIA A800, and DeepSeek V4 Flash on Hygon K100AI DCU--demonstrate deployment repair, cross-model knowledge transfer, and portability across heterogeneous accelerator software stacks.
\end{abstract}
\vspace{0.8em}
\end{@twocolumnfalse}
]

\section{Introduction}

LLM inference systems are moving targets. On the model side, new releases frequently change attention masks, sliding-window policies, KV-cache layouts, MoE routing, activation functions, quantization metadata, and projection sharding. On the hardware side, production accelerators are not replaced whenever a new model appears. Existing GPUs and AI accelerators must continue to serve new models throughout their operational lifetime.

Modern inference frameworks provide a strong foundation. vLLM improves KV-cache management and serving throughput through PagedAttention~\cite{kwon2023vllm}. TensorRT-LLM provides optimized kernels, runtime features, and deployment support for NVIDIA GPUs~\cite{nvidia2023tensorrtllm}. However, real adaptation work remains highly dependent on expert engineers. A model that runs in an upstream framework may fail in a manufacturer-maintained fork. A kernel that supports tensor-parallel degree one may not cover the shapes produced by TP or PP. A faster microbenchmark kernel may fail to improve service throughput once dispatch overhead, memory allocation, synchronization, batching, and communication are included.

This paper asks whether LLM agents can automate the adaptation and optimization of LLM inference frameworks for heterogeneous accelerators. Recent agents can read code, edit repositories, run tests, inspect logs, and repair failures over multiple rounds~\cite{jimenez2023swebench,shinn2023reflexion,qian2023chatdev}. Inference adaptation, however, is not ordinary software repair. It is coupled to hardware execution, numerical behavior, kernel dispatch, framework integration, parallel shapes, and serving objectives. Therefore, an LLM agent must be embedded in a closed execution loop.

We propose \sys, an agentic system for model-hardware co-adaptation on top of existing LLM inference frameworks. The research boundary is practical: production frameworks already contain continuous batching, KV-cache management, distributed execution, serving APIs, and deployment tools. \sys performs adaptation inside these real serving paths, then continues into end-to-end performance optimization once the model can serve correctly.

This paper makes four contributions:
\begin{itemize}
    \item We formulate LLM inference adaptation as route search over a costed execution-adaptation graph that connects model semantics, framework implementations, kernel choices, hardware capabilities, parallel shapes, and serving objectives.
    \item We separate porting from optimization: porting restores executable routes when existing paths are missing or blocked, while optimization reduces the cost of reachable routes under real serving workloads.
    \item We design \sys, an LLM-agent system that constructs the graph during execution, edits it through adapter, runtime, kernel, and configuration patches, and validates each accepted change with staged build, numerical, service, and profiling evidence.
    \item We instantiate the model with three real adaptation episodes on NVIDIA A800 and Hygon K100AI DCU, covering deployment repair, cross-model knowledge transfer, and heterogeneous accelerator portability.
\end{itemize}

\begin{figure}[t]
    \centering
    \includegraphics[width=0.98\linewidth]{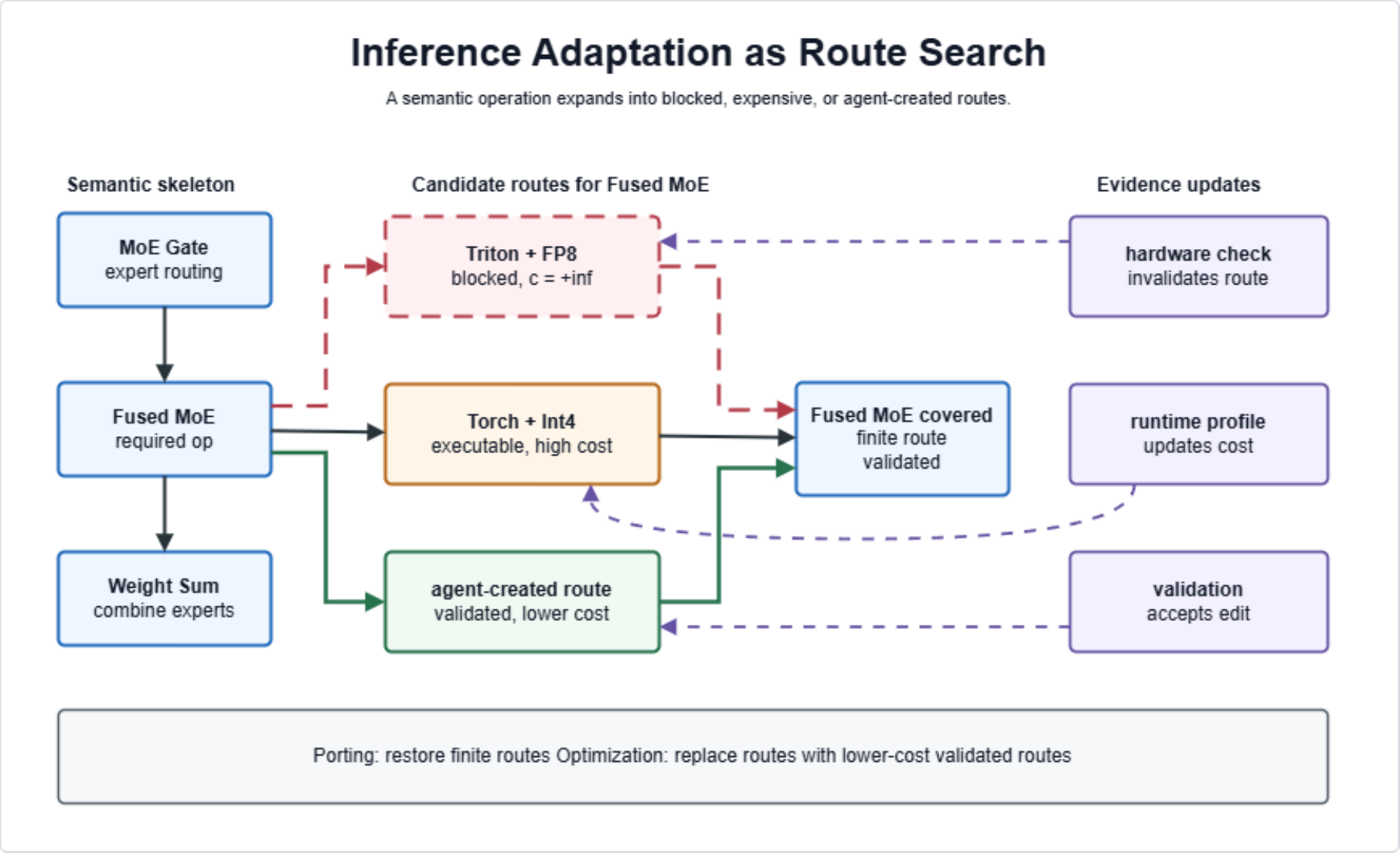}
    \caption{Early intuition of the route-search formulation. The model provides the semantic skeleton MoE Gate, Fused MoE, and Weight Sum, while Fused MoE expands into alternatives such as Triton/FP8, Torch/Int4, or an agent-created route. Hardware checks, runtime logs, validation, and profiles update route status and cost.}
    \label{fig:route-search-problem}
\end{figure}

\section{Motivation}

\subsection{The Adaptation Gap}

LLM inference software changes along several axes. First, model architectures change rapidly. A framework that supports one LLaMA-like family may fail on a model with a different attention policy, MoE layout, KV-cache organization, or quantization format. Second, hardware fleets are heterogeneous. Accelerator generations differ in memory capacity, on-chip storage, matrix units, communication topology, data type support, and compiler behavior. Third, hardware card manufacturers often maintain framework forks. These forks may diverge from upstream in model registration, operator dispatch, graph passes, kernel ABIs, communication libraries, and profiling interfaces. Fourth, parallel execution changes operator shapes. TP changes projection matrices and attention heads, PP changes activation boundaries, and expert or sequence parallelism further shifts the shape distribution seen by kernels.

Thus, making a model usable is rarely a configuration-only task. Engineers must identify semantic mismatches, implement adapters, map unsupported operators to existing libraries, write new kernels when necessary, validate numerical behavior, and reintegrate local optimizations into the full serving path. Repeating this work across models, hardware targets, and framework forks creates long adaptation cycles and high validation complexity.

\subsection{Why Existing Automation Is Insufficient}

Deep learning compilers and auto-tuning systems have long pursued performance portability. TVM provides graph-level and operator-level optimization across hardware backends~\cite{chen2018tvm}. Ansor searches a large tensor-program space with a learned cost model~\cite{zheng2020ansor}. Hidet enriches tensor-program scheduling through task mapping~\cite{ding2022hidet}. Triton lowers the barrier to writing high-performance GPU kernels~\cite{tillet2019triton,openai2021triton}. These systems are important tools for \sys, but they typically assume that the computational semantics and integration context are already well defined.

Real LLM inference adaptation often starts in a less orderly state. Model documentation may be incomplete. Framework forks may lag behind upstream. Compile-time and runtime failures may interact. Missing operators and numerical deviations may appear together. Shapes produced by TP or PP may be outside existing kernel coverage. LLM agents are useful because they can read code and logs, reason about model semantics, generate glue code, and revise hypotheses after failures. But free-form code generation is not enough. The agent must be placed inside a loop of build, execution, validation, profiling, and rollback.

\section{Problem Definition and Theory}

\subsection{Adaptation as Route Search}

We define an adaptation task as a tuple
\[
\mathcal{T} = (M, F, H, P, W, S, R),
\]
where $M$ is the model specification and checkpoint family, $F$ is the target inference framework or manufacturer fork, $H$ is the hardware and software stack, $P$ is the parallel execution plan, $W$ is the serving workload distribution, $S$ denotes correctness, deployment, and service constraints, and $R$ is the available resource budget during adaptation.

The model provides a semantic computation graph: attention, normalization, MoE routing, quantization, projection, KV-cache update, and decode steps must occur with fixed semantics. The target framework and hardware expand each semantic node into candidate execution routes. For one semantic operation, the framework may expose a native optimized kernel, a vendor library path, a Triton fallback, a PyTorch fallback, or no implementation at all. Hardware capabilities and runtime conditions then determine whether each candidate is valid, blocked, or merely expensive.

We model this state as a costed execution-adaptation graph
\[
G=(V,E,\tau,\sigma,C,\mathcal{E}),
\]
where $V$ contains model requirements, framework modules, dispatch rules, kernel implementations, fallback paths, hardware capabilities, patch actions, and execution evidence; $E$ records relations among them; $\tau$ gives each node and edge an engineering role; $\sigma$ records status such as \emph{unknown}, \emph{validated}, \emph{blocked}, or \emph{degraded}; $C$ denotes a family of costs for route reachability, serving performance, and search resources; and $\mathcal{E}$ attaches logs, tests, profiles, and accepted measurements. A route blocked by hardware or runtime conditions has infinite reachability cost. A fallback route can be reachable, yet still too expensive for the serving objective.

Under this view, porting and optimization are two phases of the same search problem. Porting restores reachability: the agent must find at least one executable route for every required semantic operation. Optimization reduces cost: once a route exists, the agent replaces expensive edges, changes dispatch conditions, or creates new kernel paths that reduce latency and memory pressure, or improve throughput and goodput under $W$. The classical shortest-path view~\cite{dijkstra1959note,hart1968formal} is useful as intuition, but a complete LLM serving path is more naturally an AND-OR structure~\cite{martelli1973additive}: all required semantic components must be covered, while each component may choose among alternative implementations. Thus, the target object is not a single linear chain, but a low-cost executable subgraph that covers the model's semantic requirements.

\paragraph{Route cost definition.}
The graph carries three distinct notions of cost. The reachability cost $c_{\mathrm{reach}}$ determines whether a route can be used at all. Routes proven invalid by compilation errors, runtime exceptions, hardware capability mismatches, or numerical failures have $c_{\mathrm{reach}}=+\infty$; validated routes have finite reachability cost; unknown routes are not blocked, but require code search, builds, minimal execution, or golden cases to determine their status. The serving cost $c_{\mathrm{serving}}$ compares validated routes under workload $W$ and serving objective $O$. The search cost $c_{\mathrm{search}}$ captures the resources consumed by an agent action, such as tokens, builds, device runs, and elapsed time. Search cost is used by the resource-aware planner, but it is not added to serving cost.

Serving cost is defined only for executable routes or candidate executable subgraphs. Let $p$ be a candidate route for a semantic operation, and let $x_j(p;W)$ denote metrics such as TTFT, TPOT, peak memory, negative throughput, or negative goodput after they have been converted to the same lower-is-better direction. We write
\[
\begin{aligned}
c_{\mathrm{serving}}(p;W,O)
&=\sum_j \omega_j \hat{x}_j(p;W),\\
\sum_j \omega_j&=1,\qquad \omega_j\ge 0,
\end{aligned}
\]
where $\hat{x}_j$ is a dimensionless normalized metric and the weights $\omega_j$ are determined by the serving objective $O$. For example, a decode-oriented objective can assign higher weight to TPOT while treating TTFT as a guard metric; a prefill-oriented objective can assign higher weight to TTFT. We do not assume that end-to-end serving cost is a linear sum of local edge costs. Local graph costs are used to rank candidates, localize bottlenecks, and explain decisions; final acceptance is still determined by measured end-to-end service metrics in the full inference framework. Descriptive and evidence nodes do not directly contribute to serving cost; executable kernel, runtime, data-movement, and communication routes do.

\subsection{Patch Space}

The system produces a patch set
\[
\Delta =
\Delta_{\mathrm{adapter}} \cup
\Delta_{\mathrm{runtime}} \cup
\Delta_{\mathrm{kernel}} \cup
\Delta_{\mathrm{config}},
\]
where each component has a distinct role. $\Delta_{\mathrm{adapter}}$ modifies the model-facing framework layer, such as model registration, configuration parsing, weight-name mapping, tensor-layout conversion, and shape guards. $\Delta_{\mathrm{runtime}}$ modifies execution routing, including dispatch guards, backend selection, fallback policy, memory allocation behavior, TP/PP integration, and KV-cache access paths. $\Delta_{\mathrm{kernel}}$ adds or tunes operator implementations, including Triton kernels, HIP/CUDA kernels, inline-assembly paths, and bindings to vendor libraries. $\Delta_{\mathrm{config}}$ modifies deployment and validation configuration, including parallel plans, dtype policy, build flags, feature switches, and workload-specific serving parameters. These four parts are separated because a successful adaptation rarely lives in one file: it changes how the framework understands the model, how it chooses a route, how critical operators execute, and how the resulting service is launched and tested.

An agent action edits the graph. It may add a missing adapter edge, mark an incompatible fast path as blocked, introduce a guarded fallback, replace a high-cost kernel edge, or attach new evidence that changes a route from unknown to validated. This connects theory to execution: the graph is not a retrospective drawing, but the control state that determines the next search, patch, and validation step.

\subsection{Adaptation Invariants}

\sys constrains agent actions through four invariants.

\paragraph{Model Semantics Invariant.}
The model is fixed. Patches may modify adapters, runtime dispatch, kernels, and configuration, but must not modify checkpoint weights, model structure, or mathematical semantics:
\[
\mathrm{Semantics}(M) = \mathrm{Semantics}(M, \delta).
\]
This separates inference adaptation from distillation, pruning, or retraining.

\paragraph{Coverage Invariant.}
Every model requirement must be covered by an executable and validated route in $G$:
\[
\begin{aligned}
\forall v \in V_M,\quad &\exists\; \mathcal{P}_v \subseteq G \\
&\mathrm{s.t.}\quad
\mathrm{Reachable}(\mathcal{P}_v)\land
\mathrm{Validated}(\mathcal{P}_v).
\end{aligned}
\]
The route may pass through a framework component, dispatch guard, kernel or fallback implementation, hardware capability node, and validation evidence. If no such route exists, the adaptation remains incomplete.

\paragraph{Evidence Invariant.}
Every accepted patch must be tied to both failure evidence and acceptance evidence:
\[
\delta_i \in \Delta \Rightarrow (e^-_i, e^+_i) \in E_{\mathrm{exec}}.
\]
Failure evidence includes compilation errors, runtime exceptions, out-of-memory events, numerical deviations, and profiling bottlenecks. Acceptance evidence includes successful builds, golden-case agreement, stable serving behavior, lower TTFT or TPOT, and improved throughput.

\paragraph{End-to-End Benefit Invariant.}
Kernel optimization must be accepted in the complete serving framework. Let $K'$ be a candidate kernel, let $F'=F\oplus K'$, and let $Q(W)$ be a service-quality vector containing TTFT, TPOT, throughput, and goodput. These metrics do not have to move monotonically in the same direction: prefill-oriented changes and decode-oriented changes may trade off TTFT, TPOT, and throughput. We therefore evaluate an optimization against a declared serving objective $O$ and a set of guard conditions $S$. Lower-is-better latency metrics directly contribute to serving cost, while throughput and goodput are converted to same-direction cost terms, such as negative throughput or relative throughput loss. The optimization is accepted only when correctness holds, the serving cost for the primary objective decreases, and monitored metrics remain within the guard conditions:
\[
\begin{aligned}
\mathrm{Accept}(K') \Leftrightarrow\;&
\mathrm{Correct}(F',H,P) \\
&\land\ c_{\mathrm{serving}}(F';W,O)
< c_{\mathrm{serving}}(F;W,O) \\
&\land\ \mathrm{Guard}_S(Q',Q),
\end{aligned}
\]
where $Q'=Q(F',H,P,W)$ and $Q=Q(F,H,P,W)$. This distinguishes \sys from single-kernel microbenchmark tuning: a faster kernel is useful only when it improves the selected service objective after framework dispatch, batching, memory allocation, synchronization, and communication are included.

\subsection{Graph State and Search History}

The execution-adaptation graph represents the current engineering state; the search tree represents the attempt history. This distinction is important. Recent agent workflow and harness self-evolution systems commonly use trees or archives to organize candidate attempts, selection, rollback, and refinement~\cite{zhou2023lats,zhang2024aflow,hu2024adas}. That is the right structure for recording how an agent explores. It is not the right structure for representing an inference stack, because hardware capabilities, dtype rules, fallback paths, dispatch guards, and validation evidence are shared by many model operations and kernels.

\sys therefore stores graph snapshots or graph deltas inside a bounded search process. A search node is
\[
n_i=(G_i,\Delta_i,E_i,B_i),
\]
where $G_i$ is the current graph, $\Delta_i$ is the applied patch set, $E_i$ is accumulated execution evidence, and $B_i$ is the remaining resource budget. Executing an action expands the search tree and updates the graph:
\[
n_j=\mathrm{Expand}(n_i,a_i), \qquad
G_{j}=\mathrm{UpdateGraph}(G_i,a_i,e_i).
\]
This resembles incremental replanning~\cite{koenig2004lifelong}: each build, run, or profile result changes edge status and cost, so the next search step operates on a more accurate graph. We do not require an admissible heuristic or claim globally optimal search. The role of the graph is to make the agent's local choices explicit: which route is blocked, which fallback is expensive, which hardware constraint is shared, and which patch can create or lower-cost a route.

\section{\sys Design}

\begin{figure*}[t]
    \centering
    \includegraphics[width=0.98\linewidth]{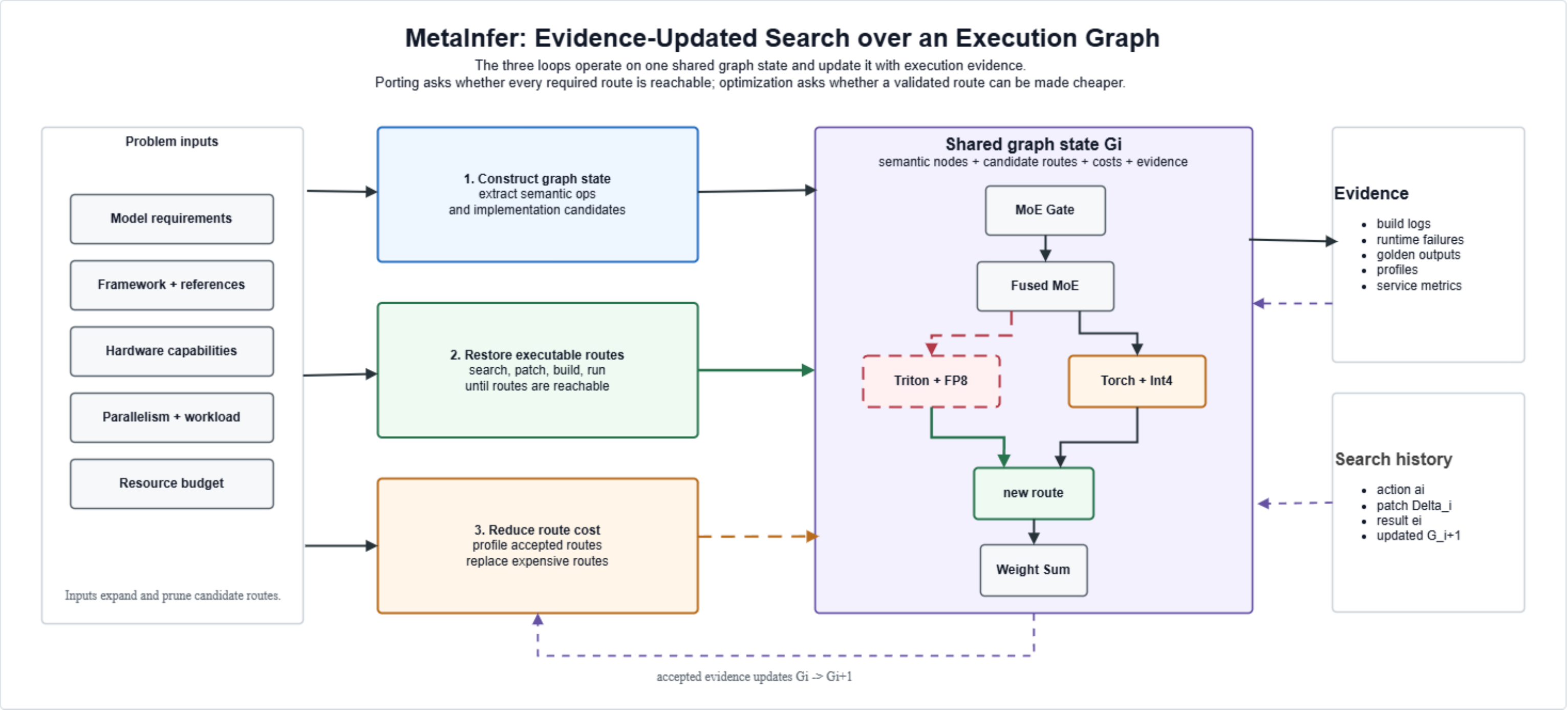}
    \caption{Three-loop architecture of \sys. The system first constructs the execution-adaptation graph, then restores executable routes, and finally reduces route cost through end-to-end performance adaptation. The loops share graph state and execution evidence so that patch search, validation, and optimization remain tied to the same traceable state.}
    \label{fig:overview}
\end{figure*}

\sys implements each adaptation episode as three connected loops. The first loop constructs graph state by parsing model configuration, checkpoint metadata, tokenizer artifacts, reference PyTorch modules, framework registries, operator dispatch tables, kernel bindings, hardware profiles, and parallel plans. The second loop restores executable routes: the agent proposes minimal graph edits as code or configuration patches, while the validation framework builds, loads, compares, and serves. The resource-aware planner expands validation from a single device to single-node multi-device and then to full service configurations. The third loop reduces route cost: after the model can serve, \sys collects end-to-end profiles, converts TTFT, TPOT, throughput, memory behavior, and operator timing into optimization hypotheses, then tunes or generates critical kernels and reintegrates them into the full framework.

\begin{table}[t]
\centering
\caption{\sys component interfaces.}
\label{tab:interfaces}
\begin{tabular}{p{0.25\linewidth}p{0.27\linewidth}p{0.32\linewidth}}
\toprule
Component & Input & Output \\
\midrule
Graph builder & Model, repository, hardware & Execution-adaptation graph \\
Migration agent & Graph, logs, tests & Adapter/runtime patches \\
Validation framework & Patches, checkpoints & Pass/fail evidence \\
Resource-aware planner & Budgets, validation cost & Staged plan \\
Profiling agent & Serving execution & Bottleneck report \\
Kernel agent & Bottleneck, shapes & Guarded dispatch patch \\
\bottomrule
\end{tabular}
\end{table}

To make early validation representative of the final serving path, \sys generates layered validation cases and reference outputs. These include sliced weights for one or a few transformer blocks, input cases that cover key paths, golden logits, intermediate tensors, routing decisions, and KV-cache states from a reference implementation. These cases and reference outputs preserve tensor layouts, data types, positional encodings, attention masks, and TP or PP shapes. Patches follow a conservative engineering discipline: small change sets, explicit tests, dispatch guards, fallbacks, regression checks, and rollback after failure.

\section{Method}

\subsection{Building the Execution-Adaptation Graph}

\sys represents model-framework-hardware constraints as a costed execution-adaptation graph. The graph starts from the model's semantic DAG, but it is expanded by repository analysis and execution evidence. A semantic node such as FP8 matrix multiplication, sparse MLA, or W8A8 GEMM is not treated as one fixed implementation. It is expanded into possible framework and kernel routes: native fast path, vendor library, Triton implementation, HIP/CUDA implementation, PyTorch fallback, or an agent-proposed route. Each route carries status, guard conditions, cost, and evidence.

\begin{figure*}[t]
    \centering
    \includegraphics[width=0.96\linewidth]{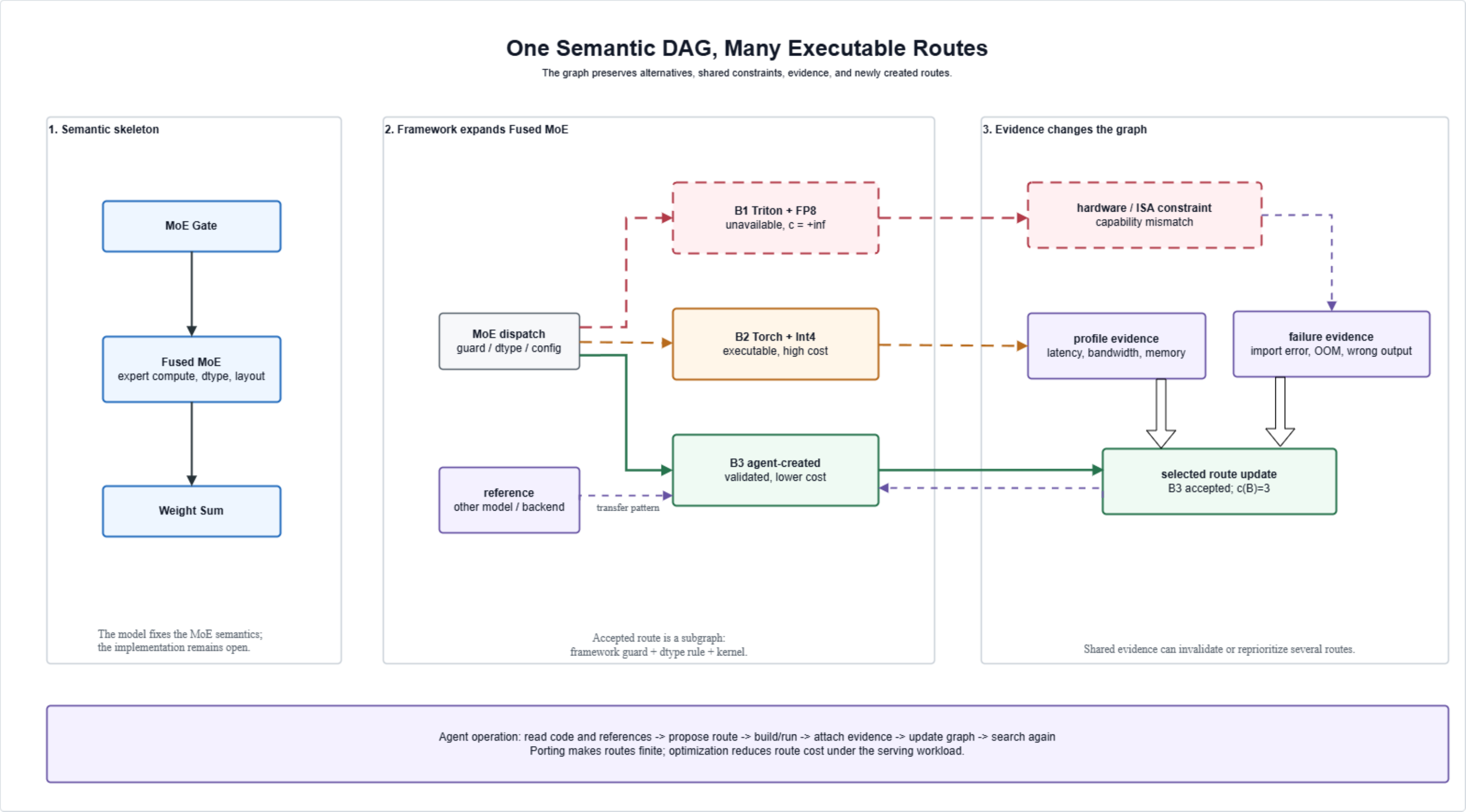}
    \caption{A costed execution-adaptation graph. The model DAG uses MoE Gate, Fused MoE, and Weight Sum as its semantic skeleton, while framework analysis expands Fused MoE into Triton, Torch, FP8, Int4, and other candidate combinations. Hardware constraints, runtime failures, and profiles update route status and cost. Porting creates at least one finite-cost route for each required operation; optimization replaces high-cost routes with lower-cost ones.}
    \label{fig:execution-adaptation-graph}
\end{figure*}

Figure~\ref{fig:execution-adaptation-graph} stays at the abstract level. It does not expose GFX928, SM80, FlashMLA, W8A8, or OOM details; those appear later in case-level subgraphs. MoE Gate, Fused MoE, and Weight Sum make the structure concrete: Fused MoE may branch into Triton, Torch, FP8, Int4, and other candidate combinations, while hardware/runtime evidence can block, validate, or make those routes expensive. This is why the graph is more useful than a chain. It preserves alternatives, shared constraints, and evidence nodes that affect more than one route.

\paragraph{Graph construction.}
The graph is built during the episode. The agent first extracts model requirements from configuration files, checkpoint metadata, model source code, and reference implementations. It then searches the target framework and reference repositories for adapters, registries, operator dispatch tables, kernel bindings, fallback rules, build flags, and tests. Hardware information comes from device queries, compilation targets, vendor documentation, existing forks, and small probing programs. Execution then turns uncertainty into evidence: build logs, import failures, minimal serving runs, golden cases, and profiles update edge status and cost.

The initial graph $G_0$ does not need to be complete. Unknown relations are kept explicitly and are resolved by build, import, minimal execution, golden cases, and profiling. Each execution step updates the graph:
\[
G_{i+1}=\mathrm{UpdateGraph}(G_i, a_i, e_i),
\]
where $a_i$ is a search, compilation, execution, patch, or profiling action, and $e_i$ is evidence from logs, exceptions, numerical comparison, or performance data. In this sense, the graph is part of the agent's working memory and control state, rather than a figure reconstructed only for exposition.

\subsection{Using the Graph for Adaptation}

The agent acts by querying and editing the graph. For functional adaptation, it asks which semantic requirements do not yet have a finite-cost validated route. If a route is missing, the agent inspects adjacent framework modules and reference implementations to produce adapter, registry, or dispatch patches. If a route is blocked by hardware capability, the agent adds guards, fallbacks, or software encoding and decoding paths. If a route executes but profiles as expensive, the agent marks it degraded and focuses optimization on neighboring kernel or runtime edges.

The same representation also supports attribution and rollback. A hardware capability node is not a dead end; it can point to multiple affected routes and to candidate patches that avoid or emulate the missing capability. A failure node is likewise not a terminal leaf; it narrows the plausible causes and prevents the agent from repeating the same patch direction. Passing validation writes acceptance evidence back to the graph. Failing validation updates status and cost, then the search tree can roll back to the nearest stable graph snapshot.

\subsection{Staged Validation}

\sys evaluates patches through expanding validation stages instead of sending every candidate to a full cluster run. Let $r_i$ be the resource configuration at stage $i$:
\[
r_1 \preceq r_2 \preceq \cdots \preceq r_k.
\]
The first stage is usually a single device or the smallest executable configuration. Later stages use single-node multi-device TP or PP, and the final stage uses the target service configuration. A patch enters an expensive stage only after passing cheaper compilation, loading, operator coverage, and numerical checks.

Low-resource validation must not hide real deployment failures. \sys therefore creates lightweight objects that preserve deployment-relevant structure: layer or block weight slices, golden cases for prefill and decode, long-context and batching patterns, TP or PP shapes, reference logits, intermediate tensors, routing results, and KV-cache states. Communication and parallelism issues that cannot be reproduced on one device are tested with the smallest multi-device configuration before full service pressure tests.

The inner loop is budgeted rather than open-ended. Code search, log attribution, local build repair, and single-device golden cases can receive several automatic attempts because feedback is cheap and precise. Single-node multi-device runs require a clearer hypothesis, and full serving runs are reserved for candidates that have already passed earlier stages. Policy choices that cannot be settled by execution, such as target SLOs or acceptable numerical tolerance, are treated as external constraints.

\subsection{Profiling-Guided Optimization}

After functional adaptation, \sys uses end-to-end profiles to update route serving costs and rank optimization candidates. It prioritizes bottlenecks that repeatedly appear in TTFT, TPOT, or throughput curves; candidates that can be validated on single-device or single-node stages; changes with limited code scope, low numerical risk, and guarded fallbacks; and optimizations that can be reintegrated into the full serving framework.

An optimized kernel is not accepted because it wins a microbenchmark. It must be evaluated under framework dispatch, memory allocation, synchronization, batching, communication, and KV-cache behavior. A profile can lower the estimated cost of a route, or mark an executable route as degraded, but final acceptance is determined by end-to-end service metrics under the serving objective $O$. This rule keeps the optimization target aligned with user-visible inference performance.

\subsection{Cross-Episode Knowledge Transfer}

A successful adaptation episode produces a transferable knowledge package:
\[
\mathcal{K}(E)=
\{C_{\mathrm{hw}}, D_{\mathrm{dispatch}}, R_{\mathrm{dtype}},
S_{\mathrm{shape}}, V_{\mathrm{test}}, P_{\mathrm{fallback}}\},
\]
where $C_{\mathrm{hw}}$ is the hardware capability matrix, $D_{\mathrm{dispatch}}$ is the dispatch rule set, $R_{\mathrm{dtype}}$ captures data-type rewriting and software encoding or decoding, $S_{\mathrm{shape}}$ records critical shape constraints, $V_{\mathrm{test}}$ contains reusable validation cases and reference outputs, and $P_{\mathrm{fallback}}$ records guarded fallback patterns.

Knowledge transfer is not patch copying. Given a source episode $E_s$ and target task $\mathcal{T}_t$, \sys initializes the target graph as
\[
G^t_0 = \mathrm{Initialize}(\mathcal{T}_t, \mathcal{K}(E_s)).
\]
Source knowledge proposes nodes, edges, guards, and validation cases for the target graph. It marks which hardware capabilities cannot be assumed, which fast paths need guards, which data types need software handling, and which shapes must enter golden cases. The transferred object is a way to construct and validate routes, not an implementation transplanted line by line.

\section{Case Studies}

We organize evaluation as real adaptation episodes. This presentation matches the nature of the task: each episode spans model understanding, framework integration, kernel work, builds, device execution, profiling, and end-to-end reintegration. The cases use NVIDIA A800 as the main narrative platform and Hygon K100AI DCU as a portability case across a different accelerator software stack. C1 repairs DeepSeek V4 Flash on NVIDIA A800 by replacing a memory-heavy FP8 conversion route. C2 transfers Ampere adaptation knowledge from DeepSeek V4 Flash to GLM 5.3 Flash. C3 adapts and optimizes DeepSeek V4 Flash on Hygon K100AI DCU across a manufacturer fork, upstream SGLang, and a non-CUDA runtime.

\begin{figure*}[t]
    \centering
    \includegraphics[width=0.98\linewidth]{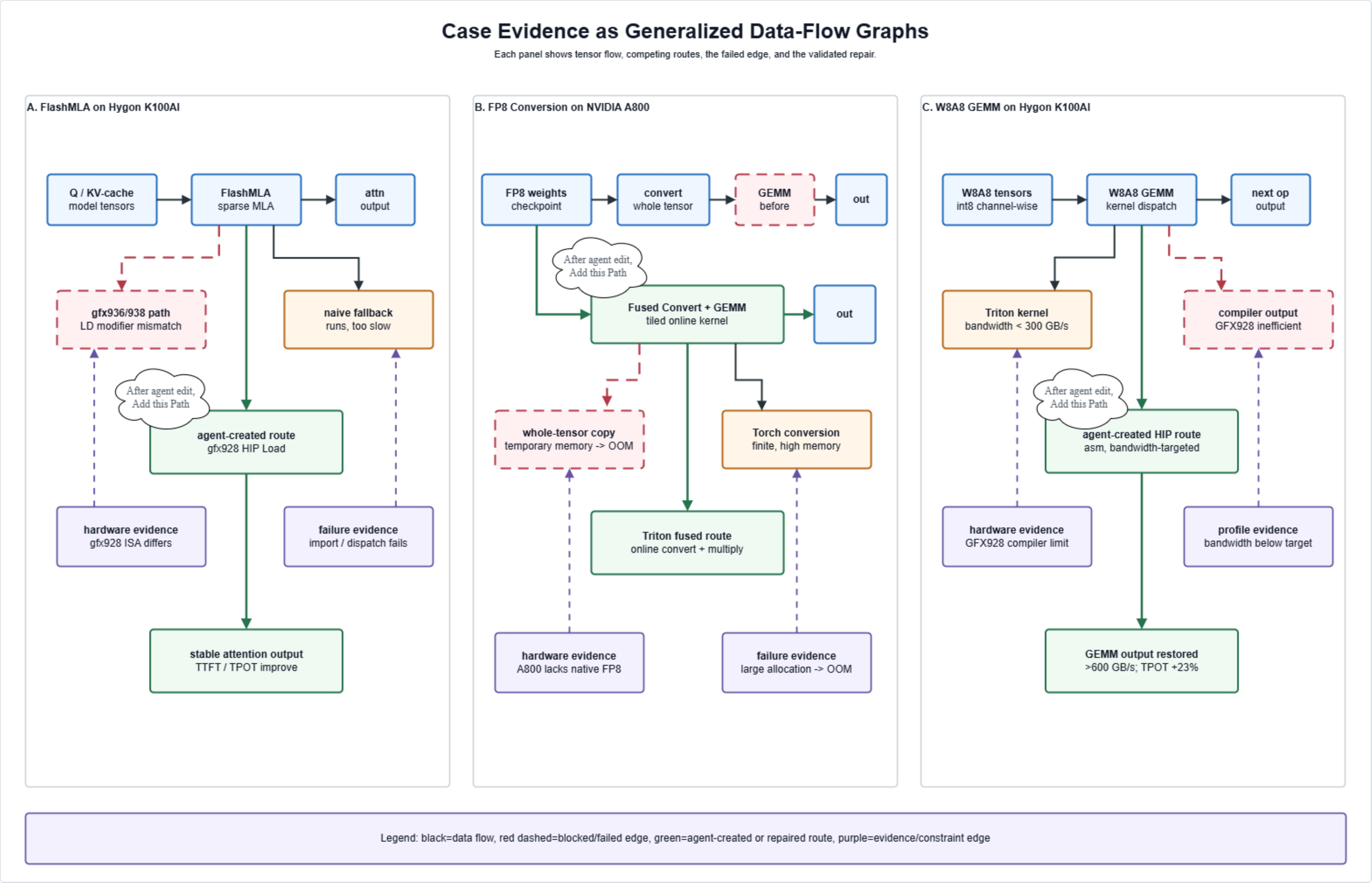}
    \caption{Generalized data-flow graphs for the three cases. Each panel shows the concrete semantic tensor flow, alternative implementations, the broken or inefficient edge, hardware or runtime evidence, and the end-to-end path restored by an agent-created or replacement route.}
    \label{fig:case-subgraphs}
\end{figure*}

\subsection{DeepSeek V4 Flash on NVIDIA A800}

The first case adapts DeepSeek V4 Flash on NVIDIA A800 using the public Ampere reference implementation at \url{https://github.com/Hakureirm/dsv4-on-ampere}. The deployment exposed a memory failure caused by data-type support mismatch. A800 is an Ampere/SM80 device and does not support the native FP8 compute path available on newer hardware. The reference implementation performed a naive whole-tensor type conversion before multiplication. Under long-context and large-tensor paths, this created a large temporary tensor and caused out-of-memory failures.

The agent localized the issue from error logs and execution paths, then rewrote the operation in Triton as tiled online conversion fused with multiplication. Each tile converts only the data it needs and immediately uses it for accumulation inside the kernel, so the runtime no longer allocates a large converted copy of the original tensor. This repair changes memory behavior at the kernel level rather than reducing batch size or context length.

Table~\ref{tab:a800-dsv4-results} reports the aggregate behavior over the full 45-workload grid and four representative workloads. The grid spans input lengths from 1k to 128k tokens, output lengths of 1k or 2k tokens, and concurrency from 1 to 16. The optimized version reduces TPOT on 42 of the 45 workloads, with an average reduction of 6.74\% and a median reduction of 6.49\%. TTFT is reported as a monitored metric because the optimization targets decode efficiency rather than prefill latency; the aggregate result makes this trade-off explicit instead of hiding it behind selected rows.

\begin{table}[t]
\centering
\caption{DeepSeek V4 Flash optimization results on NVIDIA A800. Gains are latency reductions relative to the baseline.}
\label{tab:a800-dsv4-results}
\scriptsize
\emph{Aggregate over 45 workloads.}\\[2pt]
\begin{tabular}{lrrrr}
\toprule
Metric & Improved & Mean & Median & Best \\
\midrule
TTFT & 3/45 & -4.84\% & -5.63\% & 43.83\% \\
TPOT & 42/45 & 6.74\% & 6.49\% & 26.52\% \\
\bottomrule
\end{tabular}

\vspace{4pt}
\emph{Representative workloads.}\\[2pt]
\resizebox{\linewidth}{!}{%
\begin{tabular}{lrr}
\toprule
Workload & TTFT baseline $\rightarrow$ optimized & TPOT baseline $\rightarrow$ optimized \\
\midrule
1k/1k, conc. 2 & 1075.89 $\rightarrow$ 1009.46 (6.17\%) & 22.61 $\rightarrow$ 20.90 (7.57\%) \\
1k/1k, conc. 4 & 3307.11 $\rightarrow$ 1857.56 (43.83\%) & 26.42 $\rightarrow$ 24.93 (5.64\%) \\
2k/1k, conc. 8 & 7062.87 $\rightarrow$ 6996.02 (0.95\%) & 36.27 $\rightarrow$ 33.92 (6.49\%) \\
1k/1k, conc. 8 & 3461.31 $\rightarrow$ 3541.13 (-2.31\%) & 34.66 $\rightarrow$ 32.98 (4.85\%) \\
\bottomrule
\end{tabular}
}
\end{table}

\subsection{GLM 5.3 Flash on NVIDIA A800}

The second case adapts GLM 5.3 Flash on NVIDIA A800. It uses two references: an official zero-day SGLang adaptation that does not support Ampere, and the DeepSeek V4 Flash Ampere reference from the first case. These references cannot be merged mechanically, because GLM 5.3 Flash has different DSA/KPool, sparse MLA, and block-FP8 MoE structures. The transferred knowledge is the method for reasoning about hardware capability and dispatch, not a copied patch.

The most representative transferred knowledge includes four patterns. First, backend selection should depend on per-operator capability matrices rather than GPU names or major versions alone. Second, the absence of native FP8 Tensor Core support on SM80 does not prevent storing FP8 encodings; FP8 values can be backed by \texttt{uint8} and decoded or encoded in Triton kernels. Third, Hopper-only or SM90-only fast paths should be excluded by dispatch guards and routed to portable Triton fallbacks. Fourth, quantization dispatch must carry numeric semantics such as FP8 E4M3 format, scale granularity, zero points, and weight layout, not merely \texttt{num\_bits=8}.

\begin{table*}[t]
\centering
\caption{Knowledge transfer decomposition for adapting GLM 5.3 Flash on NVIDIA A800. The episode reuses adaptation principles from DeepSeek V4 Flash on Ampere, while reimplementing model-specific runtime paths for GLM.}
\label{tab:glm-transfer}
\resizebox{\textwidth}{!}{%
\begin{tabular}{p{0.18\linewidth}p{0.28\linewidth}p{0.28\linewidth}p{0.18\linewidth}}
\toprule
Category & Knowledge from DeepSeek V4 on Ampere & GLM 5.3 Flash use & Validation evidence \\
\midrule
Capability guards & SM80 must avoid Hopper/SM90-only fast paths such as thread-block-cluster top-k, unavailable FlashMLA AOT binaries, and DeepGEMM paged-MQA paths. & Route SM80 sparse MLA and DSA/KPool index scoring to portable Triton fallbacks; fail early for explicitly selected incompatible backends. & 8$\times$A800 TP8 serving path reaches executable dispatch rather than late kernel failure. \\
FP8 representation & SM80 lacks native FP8 Tensor Core execution, but FP8 data can be stored as bytes and decoded in software. & Store DSA index cache with \texttt{uint8}/\texttt{int32} backing and perform E4M3 software decode/encode inside Triton kernels. & Removes FP8-type dispatch failures while preserving the FP8 checkpoint representation. \\
Indexer structure & The fused paged FP8 indexer avoids large gathered intermediates and combines address resolution, FP8 decode, scoring, and masking. & Generalize the indexer to GLM DSA/KPool, including paged decode and ragged prefill scorers. & Short-request TP8 execution validates the integrated DSA path. \\
Model-specific adaptation & DeepSeek V4 principles identify the right failure classes but do not provide GLM-specific kernels. & Add \texttt{D\_TAIL=0} sparse MLA support, \texttt{index\_kpool=4} cache handling, and block-FP8 Marlin MoE. & Stable serving confirms the transferred principles compose with GLM-specific runtime logic. \\
\bottomrule
\end{tabular}
}
\end{table*}

The case also required model-specific work. GLM sparse MLA permits \texttt{D\_TAIL=0}, which requires compile-time removal of zero-width tail paths. \texttt{index\_kpool=4} requires a pooled cache and ragged prefill scorer. Its expert weights use block-wise FP8 E4M3 rather than the MXFP4 path in the DeepSeek reference, requiring a new FP8 Marlin MoE path. The full TP8 model was validated on 8 NVIDIA A800 devices for short-request serving.

\subsection{DeepSeek V4 Flash on Hygon K100AI DCU}

The third case adapts and optimizes DeepSeek V4 Flash on Hygon K100AI DCU. It uses a manufacturer-adapted SGLang fork for the newer BW1000 card as a base, while also using upstream SGLang as a reference. The agent therefore reasons across two code bases, hardware generations, and a non-CUDA runtime stack.

Table~\ref{tab:k100-results} shows representative optimization results. Compared with the first correct-serving version, the final optimized version reduces TPOT from 112.70 ms to 42.57 ms under 128k context and single concurrency, a 2.65$\times$ speedup. Under 32k context and 8 concurrency, TPOT improves from 195.96 ms to 108.56 ms, a 1.81$\times$ speedup. The final version serves stably and reaches 0.6616 on GPQA-Diamond and 0.9553 on GSM8k.

\begin{center}
\refstepcounter{table}
\label{tab:k100-results}
\small
Table~\thetable: Representative DeepSeek V4 Flash optimization results on Hygon K100AI DCU.
\vspace{2pt}

\resizebox{\linewidth}{!}{%
\begin{tabular}{lrrrr}
\toprule
Workload & TTFT first $\rightarrow$ final & TTFT speedup & TPOT first $\rightarrow$ final & TPOT speedup \\
\midrule
128k context, 1 concurrency & 187364.83 $\rightarrow$ 140117.72 & 1.34$\times$ & 112.70 $\rightarrow$ 42.57 & 2.65$\times$ \\
32k context, 8 concurrency & 267941.18 $\rightarrow$ 204487.45 & 1.31$\times$ & 195.96 $\rightarrow$ 108.56 & 1.81$\times$ \\
\bottomrule
\end{tabular}
}
\end{center}

\section{Discussion and Future Work}

\sys focuses on automatic adaptation and optimization inside existing inference frameworks. This scope is deliberate. Production frameworks already contain complex serving mechanisms, and adapting a new model inside them directly targets real deployment paths. The execution-adaptation graph also provides a basis for more specialized future systems where agents maintain model- and accelerator-specific code paths as reusable route libraries.

The next step is to build a dedicated benchmark for agentic inference adaptation. Each task should include a model-framework-hardware tuple, reference implementations, failure logs, hardware capability constraints, golden cases, resource budgets, and serving workloads. Such a benchmark would support systematic comparison among general LLM coding assistants, function-only adaptation, kernel-only tuning, and the full \sys loop. It would also enable finer measurement of failure attribution, patch survival, KV-cache behavior, low-level utilization, and cross-hardware generalization.

\section{Related Work}

\paragraph{LLM inference systems.}
LLM serving frameworks optimize batching, KV-cache management, scheduling, and kernel execution. vLLM proposes PagedAttention for efficient KV-cache management and higher serving throughput~\cite{kwon2023vllm}. TensorRT-LLM provides optimized components for NVIDIA GPU inference~\cite{nvidia2023tensorrtllm}. DeepSpeed-Inference and related systems study model parallelism, memory optimization, and high-throughput generation~\cite{aminabadi2022deepspeed,holmes2024deepspeedfastgen}. \sys is complementary: it does not replace inference frameworks, but automates model adaptation and optimization inside them.

\paragraph{Deep learning compilers and auto-tuning.}
TVM, Ansor, Hidet, Triton, and FlashAttention demonstrate the value of compiler abstraction, schedule search, task mapping, programmable kernels, and IO-aware algorithms~\cite{chen2018tvm,zheng2020ansor,ding2022hidet,tillet2019triton,openai2021triton,dao2022flashattention}. \sys can use these capabilities, but its unit of automation is a full adaptation episode: identify missing support, patch framework code, validate correctness, and accept only optimizations that improve the end-to-end serving path.

\paragraph{LLM agents for software engineering.}
SWE-bench evaluates whether language models can resolve real GitHub issues~\cite{jimenez2023swebench}. Reflexion studies language agents that improve from verbal feedback~\cite{shinn2023reflexion}. ChatDev explores multi-agent software development workflows~\cite{qian2023chatdev}. \sys differs by binding agent actions to accelerator execution and serving performance. The target is not merely passing repository tests, but making a new model serve correctly and efficiently on a specific hardware stack.

\paragraph{Search and agentic graph reasoning.}
Shortest-path and heuristic-search algorithms provide the classic language for reasoning about routes and costs~\cite{dijkstra1959note,hart1968formal}. AND-OR graph search captures a closer structure for systems adaptation: all required components must be satisfied, while each component may choose among alternative implementations~\cite{martelli1973additive}. Incremental planning further explains why execution feedback should update state rather than restart search from scratch~\cite{koenig2004lifelong}. Recent LLM reasoning work also explores tree and graph structures for agent deliberation~\cite{yao2023tree,besta2024graph,zhou2023lats}. \sys uses these ideas at the systems boundary: the graph represents the inference stack, while the tree records the agent's attempts.

\section{Conclusion}

This paper presents \sys, an LLM-agent system for automatic model-hardware co-adaptation of LLM inference frameworks on heterogeneous AI accelerators. The core view is that inference adaptation is route search over a costed execution-adaptation graph. The agent must preserve model semantics, restore executable routes, reduce route cost under serving workloads, attach patches to execution evidence, and validate optimizations through end-to-end performance. Three real adaptation episodes on NVIDIA A800 and Hygon K100AI DCU demonstrate the practicality of this view and point toward a benchmark-driven agenda for AI-assisted systems engineering.

\bibliographystyle{plain}
\bibliography{references}

\end{document}